\documentclass[a4paper,aps,prd,superscriptaddress,showpacs]{revtex4}
\usepackage{graphicx}
\usepackage{epstopdf}
\usepackage{amsmath}
\usepackage{amsfonts}
\usepackage{amssymb}
\usepackage{latexsym}
\usepackage{hyperref}
\usepackage[english]{babel}
\usepackage[utf8x]{inputenc}
\usepackage[colorinlistoftodos]{todonotes}
\usepackage{xcolor}
\usepackage{slashed}
\usepackage{bm}  % To make greek letters in bold face {\bm \Phi}

\begin{document}
\title{An investigation on the electron's EDM from the electron's MDM}
%Centro Federal de Educação Tecnológica Celso Suckow da Fonseca
\author{N. Panza}\email{npanza@cefet-rj.br}
\affiliation{Departamento de F\'{i}sica, Centro Federal de Educa\c{c}\~{a}o Tecnol\'{o}gica Celso Suckow da Fonseca, Avenida Maracan\~{a}, 229, 20271-110 Rio de Janeiro, Rio de Janeiro, Brazil}
\author{Y.M.P. Gomes}\email{ymuller@cbpf.br}
\affiliation{Centro Brasileiro de Pesquisas F\'{i}sicas (CBPF), Rua Dr Xavier Sigaud 150, Urca, Rio de Janeiro, Brazil, CEP 22290-180}
\author{J. A. Helay\"{e}l}\email{helayel@cbpf.br}
\affiliation{Centro Brasileiro de Pesquisas F\'{i}sicas (CBPF), Rua Dr Xavier Sigaud 150, Urca, Rio de Janeiro, Brazil, CEP 22290-180}

%\author{}\email{}
%\affiliation{}

\begin{abstract}
\begin{center} \textbf{ ABSTRACT:} \end{center}
\paragraph*{}We pursue, in this contribution, an investigation of the contribution of the electron's magnetic dipole moment (MDM) to the electron's electric dipole moment (EDM) (or the charged leptons, more generally) in the framework of the Born-Infeld Electrodynamics and in a gravity background with a non-vanishing cosmological constant, where CP-symmetry is broken down. Our point of view consist in associating a non-trivial EDM  to a non-symmetry of the charge distribution of the elementary particle under consideration.  The bound on the electron's EDM published in 2014 is used to constrain the coupling parameter of the electromagnetic sector to the scalar curvature of the space-time in the case we consider gravity to be responsible for the asymmetry in the distribution. 
\end{abstract}
%\pacs{11.30.Cp, 12.60.-i, 13.40.Em}

\maketitle

%%%%%%%%%%%%%%
\section{Introduction} \label{sec_intro}

%\indent 

\paragraph*{}The interest in the electric dipole moment (EDM) of particles received a boost approximately fifty years ago, in a work by Ramsey, Purcell and Smith, in which the authors attempt to measure the neutron EDM \cite{EDMn}. Over the years that have gone by, the sensibility of this sort of experiment has received a major improvement, providing ways to investigate the frontiers of the Standard Model (SM)\cite{EDMs}, in the search for new physics and in the task of constraining parameters of new paths beyond the SM. A permanent EDM violates time reversal symmetry (T). If the CPT-theorem is invoked, the EDM signals a CP-violation. This takes place because the expected value for the EDM of a (massive) particle at rest is proportional to its spin, and spin is odd under T, but even under P-symmetry \cite{EDM1}.
 %The EDM of a particle is measured by a specific eletromagnetic form factor. In particular, for a spin 1/2 particle, the form factor decomposition of the electromagnnetic current is, 
%\begin{equation}  
%<f(p')| J_\mu(0)|f(p)> = \bar{u}(p')\Gamma_\mu(q) u(p)
%\end{equation} where
%\begin{eqnarray}\nonumber
%\Gamma_\mu(q) &=& F_1(q^2)\gamma_\mu + F_2(q^2) i \sigma_{\mu \nu}q^\nu / 2m \\\nonumber
%&~&+ F_A(q^2) (\gamma_\mu \gamma_5 q^2 - 2 m \gamma_5 q_\mu)\\
%&~&+ F_3(q^2)\sigma_{\mu \nu} \gamma_5 q^\nu / 2m 
%\end{eqnarray}
%with $q=p'-p$ and $m$ the mass of the fermion f \cite{EDM1}.
 \paragraph*{}In general, the EDM is calculated by the expression $d_f = - F_3(0) /2m $, where $F_3(q^2)$ in one of the electromagnetic form factors. For fermionic charges, this EDM corresponds to an effective action term $\mathcal{L}_f =- \frac{i}{2} d_f \bar{\psi}\sigma_{\mu \nu}\gamma_5 \psi F^{\mu \nu} ~~~$, whose non-relativistic regime reads as  $L_I = - H_I = d_f \vec{S}\cdot \vec{E}$. This Hamiltonian term represents the interaction energy of an EDM with an external electric field. In CP-violating renormalizable theories, this interaction flips fermion's chirality and is not invariant under the SU(2) electroweak symmetry. So, as a consequence, a non-zero value of $d_f$ yields, besides CP-violation, an explicit breaking of the eletroweak symmetry; nevertheless, as we know, the latter is spontaneously broken by the Higgs doublet. The chirality flip necessary for a non-zero EDM interaction originates from the fermion mass, which, in turn, arises from the Yukawa coupling to the Higgs doublet after spontaneous symmetry breakdown takes place \cite{EDM1}.
\paragraph*{}In gauge theories, CP-invariance could be spontaneously or explicitly broken, for example, in the case of models that present couplings between scalar fields that are not invariant under CP. According to the Cabbibo-Kobayashi-Maskawa matrix \cite{kob}, despite the sensitivity of current experiments, not only leptons, but also of baryons, EDMs are too small to be directly measured in the near future \cite{EDM1}. But, the property of a non-vanishing EDM value for truly elementary particles is the signal of some CP-violating interaction \cite{EDM1}. 

\paragraph*{}In 2014, a paper published by Science \cite{sci}, reporting an experiment with polar molecules of thorium monoxide, presents the more accurate measurement of an upper-bound for the the electron's EDM, namely:

\begin{equation}
d_e = (-2.1\pm3.7\pm 2.5) \times 10^{-29} e \cdot cm
\end{equation}
\paragraph*{}which corresponds to the upper bound $|d_e|< 8.7 \times 10^{-29} e \cdot cm$. Thus, considering that the SM (4-loop) radiative corrections contribute to  $d_e \approx 10^{- 38} e \cdot cm$, which is far below the 2014 upper-bound, this turns into a fundamental question and may constitute a very good opportunity to investigate some new physics beyond the SM. The 2014 result \cite{sci} shows that, at a scale of the order of $10^{-29}$ cm, which is only four orders of magnitude below the Planck length, gravitational effects may become relevant and could contribute to distort asymmetry the electron charge distribution around its spin direction, contributing thereby to a non-trivial EDM.

\paragraph*{}The theoretical motivation to consider supersymmetry (SUSY) in Particle Physics is that one can justify the stability of the hierarchy of the electroweak, the grand-unification and the Planck scales. In the framework of the SM, CP-violation is located in the quark sector, where the Cabibbo-Kobayashi-Maskawa phase appears in the mixing-flavor matrix. Analogue phases can be introduced by a lepton-mixing matrix and by the QCD $\theta$ term. On the other hand, SUSY may bring about new CP-violation terms among the parameters of the superpotential and in the SUSY soft-breaking terms. Differently from the CKM-matrix in the SM, CP-violation in a scenario with SUSY can already contribute at the level of 1-loop diagrams. We here point out a review \cite{EDM1} that discuss the most popular SUSY models, involving gauge multiplets of the gauge group $G_s = SU(3)_C \times SU(2)_L\times U(1)_Y $, 3 generations of left quiral supermultiplets of quarks, leptons and their supersymmetric partners and 2 Higgs supermultiplets. %The Lagrangian of a particularly interesting model reads as below:
%\begin{equation}
%\mathcal{L} = \mathcal{L}_0+ \mathcal{L}_W + \mathcal{L}_{soft}
%\end{equation}
%\paragraph*{}Where $\mathcal{L}_0$ stands for the kinetic terms, whereas the gauge interactions and $\mathcal{L}_W$ is described by the following superpotential:
%\begin{equation}
%- W = \hat{U}^c h_U \hat{Q} \hat{H}_1+ \hat{D}^c h_D \hat{Q} \hat{H}_2 +\hat{E}^c h_E \hat{L} \hat{H}_2 + \mu \hat{H}_1 \hat{H}_2 + h.c.
%\end{equation}
%where $\hat{Q}_i = (\hat{U}_i,\hat{D}_i )$, $\hat{L}_i=(\hat{N}_i, \hat{E}_i)$ and $\hat{E}_i = (l_L,\tilde{l}_L)$, $\hat{E}^{c}_i = (l^{c}_R,\tilde{l}^{*}_R)$ represents the respective lepton multiplets in a superfield formalation. The soft SUSY breaking term is:
%\begin{eqnarray}
%- \mathcal{L}_{soft}&=&\tilde{U}^{*}_R \xi_U \tilde{Q}_L H_1 + \tilde{D}^{*}_R \xi_D \tilde{Q}_L H_2 + \tilde{E}^{*}_R \xi_E \tilde{L}_L H_2 + \mu B H_1 H_2 \\ &+& \frac{1}{2} \sum_i {\mu_i}^2 z_i^{*} z_i +\frac{1}{2}\sum_a \tilde{m}_a \lambda_a \lambda_a + h.c.
%\end{eqnarray} 
  As the mass and mixing parameters which can induce CP-violation are not fully known until now, general results for $d_e$  are not really enlightening. Despite this fact, special restrictions can be worked out to estimate the contribution to the electron's EDM, as discussed in \cite{EDM1}.

\paragraph*{}Our paper is outlined according to the following organization: in Section II, we consider Born-Infeld Electrodynamics to investigate how the intrinsic MDM of a charged elementary particle contributes, even in the static regime, to its electric field and might yield distortion in the field lines so that a charge asymmetry could show up. Though Born-Infeld does not violate CP, the purpose of this Section is to check to which extent non-linearity might be underneath a non-symmetry in the charge distribution. Next, in Section II, we report our results based on the hypothesis that explicit CP-violation and gravity may distort electric field lines so that an asymmetry in the charge is generated as an effect of the particle intrinsic EDM. To close our work, we collect, in Section IV, our Discussions and Concluding Comments.

%%%%%%%%%%%%

\section{EDM influence on the electric field in Born-Infeld Electrodynamics:} \label{BI}

\indent 

\paragraph*{}In their original article, Born and Infeld proposed, based on very first principles, an electromagnetic theory valid at distance scales of the order of the radius of the electron charge, $\frac{e^2}{m c^2}$. To write down their action, they postulated that the action must be invariant under coordinate transformations, i.e., independent from the choice of coordinate system. With the tensor field  $a_{\mu \nu}$,  the Lagrangian is defined as follows:
\begin{equation}
\mathcal{L} \propto \sqrt{-\det{a_{\mu \nu}}}
\end{equation}
\paragraph*{}where the tensor may be split into its symmetric and anti-symmetric components, 
\begin{equation}
a_{\mu \nu} = g_{\mu \nu}+f_{\mu \nu} ~~,~~ g_{\mu \nu} = g_{\nu \mu}~~,~~f_{\mu \nu} = - f_{\nu \mu}
\end{equation}
\paragraph*{}The original approach of B-I electrodynamics, defined in a (1+3)-D Minkowski background, relies on the following conditions: it must reproduce Maxwell equations in a weak field limit and it must set an upper bound for the electric field in the static regime which, in turn, ensures a finite electrostatic energy once the singularity on the position of the charge is naturally removed\cite{born1}. This formulation borrows a direct analogy with Special Relativity, where there exists an upper limit on the particles velocity, the speed of light, c. They show that the most general Lagrangian that obeys these conditions, besides respecting Lorentz and gauge invariances, reads as follows:
\begin{equation}
\mathcal{L} = b^2 \sqrt{1-\frac{2 S}{b^2}-\frac{P^2}{b^4}} - b^2 \label{bi}
\end{equation}

\paragraph*{}where b is is the fundamental parameter of B-I theory, $S=-\frac{1}{4} F_{\mu \nu} F^{\mu \nu} = -\frac{1}{2}( E^2-B^2)$ and $P =-\frac{1}{4} F_{\mu \nu} \tilde{F}^{\mu \nu} =\vec{E}\cdot \vec{B}$, the well-known Lorentz invariants of Maxwell Theory. The equation \eqref{bi} can be rewritten in a more compact form, and with proper extension for a generic manifold take the form \cite{born1}: 
\begin{equation}
\mathcal{L} = b^2 \sqrt{-det(g_{\mu \nu}+\frac{1}{b} F_{\mu \nu})} - b^2\sqrt{-det(g_{\mu \nu})} + A_\mu J^\mu
\end{equation}
\paragraph*{} The previous Lagrangian reproduces Maxwell theory in the appropriate energy range. Based on the equation \eqref{bi}, the equation of motion for the gauge field is readily derived:

 \begin{equation} 
\partial_\mu(  \frac{\delta L}{\delta F_{\mu \nu}})=\partial_\mu\Bigg{(}\frac{(F^{\mu \nu}- \frac{1}{b^2}(\vec{E}\cdot \vec{B}) \Tilde{F}^{\mu \nu})}{  \sqrt{1-\frac{(E^2-B^2)}{b^2}-\frac{(\vec{E}\cdot \vec{B})^2}{b^4}}}\Bigg{)} = J^\nu
\end{equation}

\paragraph*{} and the following tensor can be defined:
\begin{equation}
G^{\mu \nu}=\frac{\delta L}{\delta F_{\mu \nu}}
\end{equation}
\paragraph*{}As a consequence, equation (11) can be rewriten as $\partial_\mu G^{\mu \nu} = J^\nu$, where, in components, we get to:
\begin{equation}
G^{0 i}=(\vec{D})_i=\frac{1}{ \sqrt{1-\frac{(E^2-B^2)}{b^2}-\frac{(\vec{E}\cdot \vec{B})^2}{b^4}}}(\vec{E}+\frac{1}{b^2}(\vec{E}\cdot \vec{B}) \vec{B} )_i
\end{equation}
\begin{equation}
\frac{1}{2}\varepsilon_{i j k} G^{j k}=(\vec{H})_i=\frac{1}{ \sqrt{1-\frac{(E^2-B^2)}{b^2}-\frac{(\vec{E}\cdot \vec{B})^2}{b^4}}}(\vec{B}-\frac{1}{b^2}(\vec{E}\cdot \vec{B}) \vec{E} )_i
\end{equation}
\paragraph*{}These are the constitutive relations, and they allow the interpretation that the non-linearity introduces both vacuum polarization and magnetization. One can write down the equations for H(E,B) and D(E,B) from equations for E(D,H) and B(D,H). By assuming axial symmetry for the current and spherical symmetry for charge density, the following equations hold:
\begin{equation}
Er = \frac{\sqrt{1-(\frac{H_r^2}{b^2}+ \frac{H_\theta ^2}{(b^2+D_r^2)})}}{\sqrt{1+\frac{D_r^2}{b^2}}}\Big{(}1+\frac{1}{b^2}\frac{(1+\frac{D_r^2}{b^2})H_\theta^2}{(1-(\frac{H_r^2}{b^2}+ \frac{H_\theta ^2}{(b^2+D_r^2)})}\Big{)} D_r
\end{equation}
\begin{equation}
E_\theta = \frac{1}{b^2} \frac{H_\theta H_r}{\sqrt{1-(\frac{H_r^2}{b^2}+ \frac{H_\theta ^2}{(b^2+D_r^2)})}}\frac{D_r}{\sqrt{1+\frac{D_r^2}{b^2}}}
\end{equation}

\paragraph*{}Applying the results above to a specific context, where the electromagnetic source is the electron, with its electric charge and its MDM, the solutions turn out to be the fields $\vec{D} = \frac{\alpha}{r^2}\hat{r}$ and $\vec{H} = \frac{\beta}{r^3} (3 (\hat{m}\cdot \hat{r})\hat{r}-\hat{m})$. Now, by expanding the results in powers of $b^{-2}$ and we may write that

%\begin{equation}
%E_r(r,\theta)= \alpha \frac{\sqrt{1-\frac{\beta^2}{r^6 b^2}\Big{(}4 \cos^2\theta + \frac{r^4 \sin^2 \theta}{(r^4+\frac{\alpha^2}{b^2})}\Big{)}}}{\sqrt{r^4+\frac{\alpha^2}{b^2}}}\Bigg{(} 1+ \frac{\beta^2 \sin^2 \theta (r^4 + \frac{\alpha^2}{b^2})}{b^2(r^{10} -\frac{\beta^2 r^4}{b^2}(4 \cos^2\theta + \frac{r^4 \sin^2 \theta}{(r^4+\frac{\alpha^2}{b^2})}))} \Bigg{)}
%\end{equation}

%\begin{equation}
%E_\theta (r,\theta)= \frac{2 q \beta^2}{b^2 r^3(\sqrt{r^4+\frac{\alpha^2}{b^2}})}\frac{1}{\sqrt{r^6-\frac{\beta^2}{b^2}(4 \cos^2\theta + \frac{r^4 \sin^2 \theta}{(r^4+\frac{\alpha^2}{b^2})})}}
%\end{equation}

\begin{eqnarray}
E_r(r,\theta) &=& \frac{\alpha}{r^2} - \frac{2 \alpha^3}{b^2 r^6} - \frac{\alpha\beta^2(3+5 \cos 2 \theta)}{b^2 r^8}+O(b^{-4}) \\
E_\theta(r, \theta) &=& - \frac{\alpha\beta^2\sin 2 \theta}{b^2 r^8}+O(b^{-4})
\end{eqnarray}

\paragraph*{}As we may check above, though there is a non-trivial vacuum polarization that could, in principle, contribute to a non-trivial EDM, the total polarization vanishes whenever integrated over the whole space ($\vec{P}=\vec{D}- \frac{1}{4\pi}\vec{E}$)
\begin{equation}
\delta\vec{d}_e = \int d^3 x \vec{P} = 0
\end{equation}
\paragraph*{} So, the non-linear character of B-I theory by itself is not able to induce a non-trivial EDM. CP-breaking of the underlying action must actually be a necessary requirement. Our initial expectation was that a tiny contribution to the asymmetry of the electron's charge distribution might arise from non-linearity; however, CP-symmetry is stronger than non-linearity and so no EDM contribution comes out in a Born-Infeld approach. 
\paragraph*{}By keeping in mind the importance of CP-violation for the electron's EDM, we change our scenario from a non-linear electromagnetic model to pursue an investigation where gravity effects may, in combination with the intrinsic magnetism of the charged particle, produce a distortion on the electric field lines so as to affect the symmetry of the charge distribution around the spin of the particle. We are also motivated by the fact that the the upper limit on $d_e$ displays a the length scale $10^{-29}$ cm, which is close to the Planck's length, so that gravity effects may show up with a relevant contribution.

%%%%%%%%%%%%%%%%%%%
\paragraph*{}
\section{ A Maxwell-gravity coupled model with $RF\tilde{F}$ and non-trivial EDM}

\indent 
\paragraph*{} In order to be self-contained and to set up our notation and conventions, we briefly review some
basic notions. The Einstein-Maxwell model in the presence of a cosmological term can
be written as follows:%
\begin{equation}
S_{EM}=%
%TCIMACRO{\dint }%
%BeginExpansion
{\displaystyle\int}
%EndExpansion
d^{4}x\sqrt{-g}\left[  \alpha\left(  R-2\Lambda\right)  -\frac{1}{4}F_{\mu\nu
}F^{\mu\nu}-J^{\mu}A_{\mu}\right]
\end{equation}
where $\alpha=\dfrac{1}{16\pi G}$, $G$ is the Newtonian gravitational constant
and $g$ is the determinant of the metric tensor $g_{\mu\nu}$. $R$ and
$\Lambda$ represent, respectively, the scalar curvature and cosmological constant.
As usually, $F_{\mu\nu}=\partial_{\mu}A_{\nu}-\partial_{\nu}A_{\mu}$ is the
electromagnetic field tensor and $J^{\mu\text{ }}$is the current density. It is worthy
noticing that, in our scenario, as gravity acts directly on the field lines of the
electromagnetic field, an electric dipole moment can appear in view of the distortion of these lines. Furthermore, as well-known, in particle
physics, symmetries and their eventual violations have been an important guide for building up successful models.
In this sense, the existence of an electric dipole moment of the
electron and charged leptons in general would manifest as a consequence of a more fundamental physics that underlines CP-violation. 
Based on this fact, in order to better understand the effect of gravity on the EDM, we modify the action (1) by adding up a non-minimal 
coupling  between the Maxwell field and the scalar curvature of the background geometry that represents gravity. The simplest of these models 
is given by the action%
\begin{equation}
S=%
%TCIMACRO{\dint }%
%BeginExpansion
{\displaystyle\int}
%EndExpansion
d^{4}x\sqrt{-g}\left[  \alpha\left(  R-2\Lambda\right)  -\frac{1}{4}F_{\mu\nu
}F^{\mu\nu}+\beta RF_{\mu\nu}\widetilde{F}^{\mu\nu}-J^{\mu}A_{\mu}\right]
\end{equation}
where $\beta$ is an arbitrary parameter with mass dimensional equal to (-2) and $\widetilde{F}_{\mu\nu}=(1/2)\epsilon_{\mu\nu\alpha\beta}F^{\alpha\beta
}$ is the dual electromagnetic field-strength tensor and $\epsilon_{\mu
\nu\alpha\beta}=\sqrt{-g}\varepsilon_{\mu\nu\alpha\beta}$, $\varepsilon
_{\mu\nu\alpha\beta}$ is the Levi-Civita tensor density. 
The field equations derived upon variation
with respect to the potential $A^{\mu}$ are
\begin{equation}
\frac{1}{\sqrt{-g}}\partial_{i}\left(  \sqrt{-g}g^{00}D^{i}\right)  =\rho
\end{equation}
\begin{equation}
\frac{1}{\sqrt{-g}}\partial_{i}\left(  \sqrt{-g}\epsilon^{0ji}\text{ }%
_{k}H^{k}\right)  -\frac{1}{\sqrt{-g}}\partial_{0}\left(  \sqrt{-g}g^{00}%
D^{j}\right)  =J^{j}%
\end{equation}
where $D^{i}$ and $H^{i}$ fields are defined in the sequel.

\paragraph*{} The homogeneus Maxwell equations can be written as%
\begin{equation}
\frac{1}{\sqrt{-g}}\partial_{i}\left(  \sqrt{-g}g^{00}B^{i}\right)
\end{equation}
\begin{equation}
\frac{1}{\sqrt{-g}}\partial_{0}\left(  \sqrt{-g}g^{00}Bj\right)  +\frac
{1}{\sqrt{-g}}\partial_{i}\left(  \sqrt{-g}\epsilon^{0ij}\text{ }_{k}%
E^{k}\right)  =0
\end{equation}

\paragraph*{} Throughout the paper, we adopt the following notational convention for
the components of the electric ($E_{i}$) and magnetic induction ($B_{i}$)
field%
\begin{equation}
F_{0i}=-E_{i}%
\end{equation}
\begin{equation}
\widetilde{F}_{0i}=-B_{i}%
\end{equation}
 \paragraph*{} In addition to
these definitions, the displacement vector $Di$ and the magnetic field $B_{i}$
must satisfy the constraints%
\begin{equation}
D_{i}=E_{i}-4\beta RB_{i}%
\end{equation}
\begin{equation}
H_{i}=B_{i}-4\beta RE_{i}%
\end{equation}

\paragraph*{} Obviously, if $R=0$, we recover the constitutive relations for the flat space.

\paragraph*{} Before we go further, we must specify that, besides the Lagrangian that describes
the dynamics of the fields, the space-time structure where the model is built up plays a central rôle. 
Our choice of this structure will be based on the fact that recent
cosmological observations provide a strong evidence that there is a positive
cosmological constant in our universe\cite{Eric}. Thus, the space-time is asymptotically a de
Sitter space. Furthermore, according to the standard cosmological model, the
Universe is presently dominated by a dark energy and one of the favorite
candidate for this is the cosmological constant. On the other hand, dark
matter has been postulated to explain various observations from gravitational
effects on visible matter. Similarly, we can postulate that the cosmological
constant directly affects the charge distribution of the electron in order to
induce a electrical dipole moment.

\paragraph*{} Let us start off with the of the Sitter solution in four dimensional space-time
with spherically-symmetric line element%
\begin{equation}
ds^{2}=\left(  1-\frac{\Lambda}{3}r^{2}\right)  dt^{2}-\left(  1-\frac
{\Lambda}{3}r^{2}\right)  ^{-1}dr^{2}-r^{2}\left(  d\theta^{2}+sen^{2}\theta
d\phi^{2}\right)
\end{equation}
where $\Lambda=-4R$.

\paragraph*{} In the structure of the space-time described above, let us assume that we place a point-like
charge. The charge density is given by%
\begin{equation}
\rho\left(  \overrightarrow{r}\right)  =-e\delta^{3}(\overrightarrow{r})%
\end{equation}

\paragraph*{} We can go one step further in our study by assuming the static regime and that no stationary current is present. In
this situation, equations (18-21) take the form%
\begin{equation}
\frac{1}{\sqrt{-g}}\partial_{i}\left(  \sqrt{-g}g^{00}D^{i}\right)
=-e\delta^{3}\left(  \overrightarrow{r}\right)
\end{equation}
\begin{equation}
\frac{1}{\sqrt{-g}}\partial_{i}\left(  \sqrt{-g}\epsilon^{0ji}\text{ }%
_{k}H^{k}\right)  =0
\end{equation}
\begin{equation}
\frac{1}{\sqrt{-g}}\partial_{i}\left(  \sqrt{-g}\epsilon^{0ij}\text{ }%
_{k}E^{k}\right)  =0
\end{equation}
\begin{equation}
\frac{1}{\sqrt{-g}}\partial_{i}\left(  \sqrt{-g}g^{00}B^{i}\right)  =0
\end{equation}

\paragraph*{} These equations can be rewritten in terms of the cosmological constant in the form%
\begin{equation}
\overrightarrow{E}=\overrightarrow{D}-16\beta\Lambda\overrightarrow{B}%
\end{equation}
\begin{equation}
\overrightarrow{H}=\left(  1-256\beta^{2}\Lambda^{2}\right)  \overrightarrow
{B}+16\beta\Lambda\overrightarrow{D}%
\end{equation}

\paragraph*{} As discussed in the previous Section, we are interested in the EDM of the charged 
particle. Notwithstanding, we shall be seeking solutions to the equations (28-31) that exhibit
radial and angular dependences. Thus, we propose the solutions to be expressed as the ansätze below:
\begin{equation}
\overrightarrow{H}\left(  r,\theta\right)  =H_{r}\left(  r,\theta\right)
\widehat{r}+H_{\theta}\left(  r,\theta\right)  \widehat{\theta}%
\end{equation}
\begin{equation}
\overrightarrow{B}\left(  r,\theta\right)  =B_{r}\left(  r,\theta\right)
\widehat{r}+B_{\theta}\left(  r,\theta\right)  \widehat{\theta}%
\end{equation}

\paragraph*{} The solution to eq. (27) is given by
\begin{equation}
\overrightarrow{D}=\frac{e}{r^{2}}\left(  1-\frac{\Lambda r^{2}}{3}\right)
^{1/2}\left(  -\widehat{r}\right)
\end{equation}

\paragraph*{} Inserting equation (34) and (36) into equation (33), we obtain%
\begin{equation}
H_{r}\left(  r,\theta\right)  =aB_{r}\left(  r,\theta\right)  +bD_{r}\left(
r\right)
\end{equation}
\begin{equation}
H_{\theta}\left(  r,\theta\right)  =aB_{\theta}\left(  r,\theta\right)
\end{equation}
where, for sake of simplicity, we take $a=1-256\beta^{2}\Lambda^{2}$ and
$b=16\beta\Lambda$.

 \paragraph*{} In order to solve the equations above, we adopt the method of separation of variables.
We do not reproduce below the details of the lengthy calculations. We simply quote the results of interest%
\begin{equation}
B_{r}\left(  r,\theta\right)  =\left(  1-\frac{\Lambda r^{2}}{3}\right)
^{1/2}%
%TCIMACRO{\dsum \limits_{i=0}^{\infty}}%
%BeginExpansion
{\displaystyle\sum\limits_{i=0}^{\infty}}
%EndExpansion
\frac{d}{dr}\left[  c_{l}r^{1}\text{ }_{2}F^{1}\left(  \frac{1}{2}+\frac{l}%
{2},\frac{l}{2},\frac{3}{2}+l,\frac{\Lambda r^{2}}{3}\right)  +d_{l}%
r^{-\left(  l+1\right)  }\text{ }_{2}F^{1}\left(  -\frac{1}{2}-\frac{l}%
{2},-\frac{l}{2},\frac{1}{2}-l,\frac{\Lambda r^{2}}{3}\right)  \right]
P_{l}\left(  \cos\theta\right)
\end{equation}
\begin{equation}
B_{\theta}\left(  r,\theta\right)  =%
%TCIMACRO{\dsum \limits_{l=0}^{\infty}}%
%BeginExpansion
{\displaystyle\sum\limits_{l=0}^{\infty}}
%EndExpansion
\left[  c_{l}r^{l-1}\text{ }_{2}F^{1}\left(  \frac{1}{2}+\frac{l}{2},\frac
{l}{2},\frac{3}{2}+l,\frac{\Lambda r^{2}}{3}\right)  +d_{l}r^{-\left(
l+2\right)  }\text{ }_{2}F^{1}\left(  -\frac{1}{2}-\frac{l}{2},-\frac{l}%
{2},\frac{1}{2}-l,\frac{\Lambda r^{2}}{3}\right)  \right]  \frac{dP_{l}\left(
\cos\theta\right)  }{d\theta}%
\end{equation}
where $_{2}F^{1}$ is the hypergeometric function and $e$ is the electron charge.

\paragraph*{} From the boundary condition at infinite, namely, $B\rightarrow0$, when
$r\rightarrow\infty$, we find that the only nonvanishing non-vanishing coefficient is
$d_{l}$. These coefficients are determined as in the sequel.
Here, we are taking the point of view that gravity effects will interfere in the process  of computing the influence of the $\overrightarrow{B}$-field lines on the electric field. The electron has an intrinsic magnetic moment; this magnetic moment is the source of a weak magnetic field and gravity connect the latter with the electric field of the electron charge. Since gravity is weak at this scale, we adopt the usual magnetic field, $\overrightarrow{B}$, of a dipole, without including
gravity effects in its expression. The effect of gravity shall appear whenever computing the influence of the latter on the electrostatic field. Should we include gravity effects in the expression of $\overrightarrow{B}$, we would have a second-order effect, that we are neglecting.
The magnetic induction created by a spin current distribution localized in a
small region of space and in absence of gravity is given by%
\begin{equation}
\overrightarrow{B}\left(  \overrightarrow{r}\right)  =\frac{3\overrightarrow
{r}\left(  \overrightarrow{m}\cdot\overrightarrow{r}\right)  }{r^{5}}%
-\frac{\overrightarrow{m}}{r^{3}}+\overrightarrow{m}\delta^{3}\left(
\overrightarrow{r}\right)
\end{equation}
where $\overrightarrow{m}$ is the magnetic moment of the distribution. Let us assume that the magnetic moment of the electron is oriented along
the $z$ axis. In this case, the radial and polar components of the equation (41) are given by%
\begin{equation}
B_{r}\left(  r,\theta\right)  =\frac{2m\cos\theta}{r^{3}}%
\end{equation}
\begin{equation}
B_{\theta}\left(  r,\theta\right)  =\frac{m\sin\theta}{r^{3}}%
\end{equation}

\paragraph*{} Taking $\Lambda=0$ in equations (39) and (40), and comparing term by term with equations (42) and (43), the only piece that survives is $d_{1}=m$. Finally, the components of the magnetic induction obtained from the equations (39) and (40) can be written as%
\begin{equation}
B_{r}\left(  r,\theta\right)  =-\frac{2m}{r^{3}}\left(  1-\frac{\Lambda r^{2}%
}{3}\right)  ^{1/2}\cos\theta
\end{equation}
\begin{equation}
B_{\theta}\left(  r,\theta\right)  =-\frac{m}{r^{3}}\left(  1-\frac{\Lambda
r^{2}}{3}\right)  \sin\theta
\end{equation}

\paragraph*{} Making use of the last equations along with equation (32), we
obtain that the components of the electric field are given by%
\begin{equation}
E_{r}\left(  r,\theta\right)  =\left(  1-\frac{\Lambda r^{2}}{3}\right)
^{1/2}\left(  -\frac{e}{r^{2}}+\frac{32\beta\Lambda m}{r^{3}}\cos
\theta\right)
\end{equation}
\begin{equation}
E_{\theta}\left(  r,\theta\right)  =\left(  1-\frac{\Lambda r^{2}}{3}\right)
\frac{16\beta\Lambda m}{r^{3}}\sin\theta
\end{equation}

\paragraph*{} From these equations, we obtain that the combined effect CP-violation and gravity (through the cosmological
constant) on the charge density of the particle is given by the equation%
\begin{equation}
\frac{1}{\sqrt{-\widetilde{g}}}\partial_{i}\left(  \sqrt{-\widetilde{g}}%
E^{i}\right)  =\rho
\end{equation}
where $\widetilde{g}$ is the determinant of the spatial sector of the de
Sitter metric, which leads to%
\begin{equation}
\rho\left(  r,\theta\right)  =\frac{\Lambda}{3r}\left(  e-\frac{32\beta\Lambda
m\cos\theta}{r}\right)
\end{equation}

\paragraph*{} We are now ready to find the electric dipole moment of the electron. We
perform this by using the relation%
\begin{equation}
\overrightarrow{d_{e}}=%
%TCIMACRO{\dint }%
%BeginExpansion
{\displaystyle\int}
%EndExpansion
\sqrt{-\widetilde{g}}\overrightarrow{r}\rho\left(  \overrightarrow{r}\right)
d^{3}r
\end{equation}

\paragraph*{} The result may be put in the form%
\begin{equation}
\overrightarrow{d_{e}}=\frac{64\pi\beta\Lambda m}{3}\left[  1-\sqrt
{1-\frac{\Lambda r_{0}^{2}}{3}}\right]  \left(  -\widehat{z}\right)
\end{equation}
where $r_{0}$ is the electron classic radius. It is helpful to bear in mind
that the product $\Lambda r_{0}^{2}$ is very small and thus it can be
simplified to%
\begin{equation}
\overrightarrow{d_{e}}=\frac{32\pi\beta\Lambda^{2}r_{0}^{2}}{9}\left(
-\overrightarrow{m}\right)
\end{equation}
\paragraph*{} Write in the international system of units, the last result take the form
\begin{equation} 
\overrightarrow{d_{e}}=\frac{32\pi\beta\mu_{0}\Lambda^{2}r_{0}^{2}}{9c}\left(
-\overrightarrow{m}\right)
\end{equation}

\paragraph*{} With this result, we can estimate the upper limit for the $\beta$ coupling
constant. To this purpose, we use the most recent values $\left\vert \overrightarrow{d_{e}}\right\vert \approx8,7\times10^{-29}e\cdot
cm$ \cite{sci}, $\left\vert \overrightarrow{m}\right\vert \approx9,28\times10^{-24}A
$ m$^{\text{-2}})$ \cite{blum}, $\Lambda\approx\times10^{-52}$ m$^{-2}$ \cite{stefano}, $r_{0}\approx2,8\times10^{-15}$ m,  $\mu_{0}=4\pi\times10^{-7}$T m A$^{-1}$ we obtain%
\begin{equation}
\beta\approx10^{120}~\text{C}^{2}\text{m}~\text{kg}^{-1}%
\end{equation}
\paragraph*{} In natural units, the value of the coupling constant is given by
\begin{equation}
\beta\approx10^{146}~\text{Gev}^{-2}
\end{equation}

\paragraph*{} Our study shows that the intrinsic magnetic field of a charged particle, in presence of a cosmological constant, breaks the radial symmetry of the electric field lines associated to the (charge) monopole. The MDM, in connection with the space-time curvature, is able to distort the electric field lines and to induce an asymmetric in the charge distribution which, as a result, yields a non-trivial EDM. So, we understand that gravity may interpolate between the MDM and the EDM of spinning charged particles.
\paragraph*{} However, if we take the most recent upper bound on the electron's EDM, the coupling parameter, $\beta$, as estimated above, sets up a huge length scale or, equivalently, a very tiny mass scale, as a consequence of the extremely small value of the cosmological constant. On the other hand, we should keep in mind that $\beta$ is not a universal coupling parameter; it is instead species-dependent: the different lepton and quark families exhibit each a different value for $\beta$.

\section{Discussion and Concluding Comments}

\indent 

\paragraph*{} In this contribution, we have focused our efforts to a discussion on the EDM of the electron, which, as motivated in the Introduction, is getting a great deal of momentum over the past years.
\paragraph*{} We have associated the discussion of the electron's EDM with a possible non-linearity of Electrodynamics - here described in a Born-Infeld set-up - just to understand how the intrinsic 
MDM do the electron (or a charged lepton in general) influences on the electrostatic sector. A 
magnetostatic field, as a consequence of the non-linearity, may be the source for an electrostatic
field configuration. The magnetic field lines being non-symmetric could play the rôle of a non-symmetric charge distribution, which, in turn, would correspond to a non-trivial EDM. The B-I scenario, however, does not violate CP, therefore, we should not expect an EDM contribution in this description. Nevertheless,this initial study was a pre-heating to our true goal: a possible relationship between the cosmological constant and the intrinsic magnetism of the electron and the tiny electron's EDM.
\paragraph*{} The coupling between  Electrodynamics and a gravitational background  clearly shows how a magnetostatic field may induce  a contribution to the electrostatic  sector. This is why  we have tackled this issue in the framework of BI Electrodynamics. And, actually, at it should be expected, the MDM, through its magnetostatic field, does not induce a no-symmetric charge distribution so that no EDM come out as a by-product of plain non-linearity, unless there was a CP-violating term in connection with BI. 
\paragraph*{} By considering a gravity background with cosmological constant, we follow the path of the MDM as a possible ingredient for yielding a non-symmetric configuration of the lines of the electrostatic fields generated by the electron charges. The viewpoint we have pursed here is that the intrinsic magnetic field of the electron might, through the presence of the gravitational background, distort the  radial electrostatic field lines associated to the electron charge so as to finally account for the non-symmetric charge distribution and then, to a non-vanishing EDM for the electron.
\paragraph*{} The CP-violating interaction term which gives rise to EDM, which involves the scalar curvature  of space-time (with cosmological constant), is governed by a coupling constant with mass dimension equal to (-2). Our  developments show that the latter appears linearly related to the electron's EDM. By considering the most recent upper bound on its experimental value, we get a huge value of the coupling parameter responsible for the EDM of the electron. We understand that this is consequence of the extremely small value of the cosmological constant. Our work just puts into evidence that we may establish a relationship between the cosmological constant and the non-symmetric of the electron's charge distribution, characterized by a length scale ($10^{-29}$ cm) not so far from the Planck length, which may be a motivation to associate the electron's EDM as a distortion which may find its origin in the space-time curvature. 
\paragraph*{} Adopting a different point of view to evaluate effects of gravity on the EDM, the work of Ref. \cite{andrew} calculates that the Earth's gravity alone may cause the muon's spin to precess. This is very appealing, for the precession of the spin (by virtue of gravity) responds for the asymmetry of the charge  distribution, yielding, therefore a non-trivial EDM, in this case, the muon's EDM.
\paragraph*{} The initial results we report in the present paper encourage us to go further and endeavour another line of studies: fermions couple to space-time torsion and, then, we should account for quantum effects of dynamical torsion in the calculation of the lepton's EDM. Extensions of gravity with higher powers of curvature and dynamical torsion emerge whenever we take field-theory limit of string models. So, a 1-loop computation of the EDM in this class of models would clarify to which extent quantum gravity effects actually contribute to the EDMs of truly elementary particles. This is work in progress and our results on that shall be report in a forthcoming paper.

\begin{center}
\textbf{Acknowledgment:}
\end{center}

The authors are deeply indebted to Dr. H. Rodrigues and Dr. R. C. Paschoal for your pertinent suggestions on the original manuscript of this paper. Y. M. P. Gomes  is grateful to CNPq for his graduate fellowship.

\end{document}